\documentclass[prl,twocolumn,showpacs,preprintnumbers,amsfonts,a4paper]{revtex4-1}

\usepackage[sort&compress]{natbib}
\usepackage{graphicx} 
\usepackage{epstopdf}
\usepackage{dcolumn}	 
\usepackage{bm}		 
\usepackage{amssymb,amsmath}
\usepackage{latexsym}
\usepackage{color}
\usepackage[breaklinks,colorlinks,bookmarks=false,citecolor=blue,linkcolor=red,urlcolor=blue]{hyperref}

\graphicspath{{./}, {Figures/}}

\newcommand{\bea}{\begin{eqnarray}}
\newcommand{\eea}{\end{eqnarray}}

\newcommand{\be}{\begin{equation}}
\newcommand{\ee}{\end{equation}}

\newcommand{\im}{{\mathrm{i}}}
\newcommand{\Tr}{{\mathrm{Tr}}}

\newcommand{\phimax}{\Phi_{\rm max}}
\newcommand{\Umax}{U_{\rm max}}

\begin{document}

\title{Adiabatic Tracking of a State: a New Route to Nonequilibrium Physics}

\author{M.~Moliner}
\affiliation{IPCMS (UMR 7504) and ISIS (UMR 7006), Universit\'e de Strasbourg et CNRS, 67000 Strasbourg, France}
\affiliation{Laboratoire de Physique Th\'eorique et Mod\'elisation, CNRS UMR 8089, Universit\'e de Cergy-Pontoise, 
Site de Saint-Martin, 2 Avenue Adolphe Chauvin, 95302 Cergy-Pontoise Cedex, France}

\author{P.~Schmitteckert}
\affiliation{Institute for Nanotechnology, Karlsruhe Institute for Technology, 76344 Eggenstein-Leopoldshafen, Germany}
\affiliation{Center for Functional Nanostructures, Karlsruhe Institute for Technology, 76131 Karlsruhe, Germany}

\date{\today}

\begin{abstract}
We present a novel numerical approach to track the response of a 
quantum system to an external perturbation that is progressively switched on.
The method is applied, within the framework of the density matrix
renormalization group technique, to track current-carrying states of 
interacting fermions in one dimension and in the presence of an Aharonov-Bohm magnetic flux. 
This protocol allows us to access highly excited states.
We also discuss the connection with the entanglement entropy of these excited states.
\end{abstract}

\pacs{71.10.-w, 03.67.Mn, 75.40.Mg}


\maketitle


Long-lived excited states have attracted a large interest in relation with exotic states of matter such as topological excitations.\cite{Alet2006} 
Cold atomic gases provide a new playground to study excited states due to the tunability of the interactions and the decoupling from the environment \cite{Bloch_review, *Review_quenches} which allow, for example, the stabilization of gases of excited states.\cite{Haller_2009} 
The absence of thermalization observed experimentally in quasi one dimensional (1D) systems after a brutal variation of one parameter (quantum quench) has motivated intense research about their stationary behavior.\cite{Rigol_2007, *Rigol_2008} 
New approaches to determine the physical properties of excited states are thus required to compute observables after a quench or to start a quench dynamics starting from an excited state.\cite{Deng_2011}

\begin{figure}[hb]
\includegraphics[width=0.47\textwidth,clip]{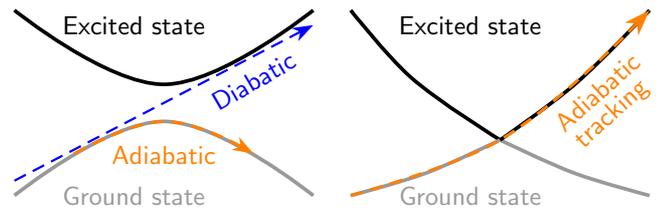} 
\caption{\label{fig:cartoon}Left panel: Landau-Zener effect. At an avoided level crossing, a diabatic evolution allows a transition to an excited state while in an adiabatic evolution the system remains in its GS. Right panel: Without a gap, the adiabatic tracking allows reaching excited states.}
\end{figure}

In this work, we present a general numerical method to track the response of a system to an external perturbation that is slowly switched on. 
We designate it as \emph{adiabatic tracking} in reference to the Landau-Zener effect \cite{Landau_1932, *Zener_1932, *Landau-Zener}, 
which considers transitions at an avoided crossing between two energy levels in a time-dependent Hamiltonian (see Fig.~\ref{fig:cartoon}).
Our technique aims at the same goal as the counterdiabatic approach. \cite{Demirplak_2003, *Demirplak_2005, *Demirplak_2008, *Berry_2009, *DelCampo_2013} 
There, one searches for additional terms to obtain an adiabatic state evolution from an unitary time evolution under the extended Hamiltonian. 
In contrast we can directly work with the Hamiltonian of interest.
An experimental realization of our technique would therefore be subject to the Kibble-Zurek mechanism.\cite{Dziarmaga_2010} 
The latter states that no matter how slowly a system (quantum or classical) is driven across a continuous phase transition, one cannot follow adiabatically its instantaneous ground state (GS) near the critical point in the thermodynamic limit due to the vanishing of the energy gap. 
This mechanism does not explicitly apply to the technique described below, as we follow eigenstates based on a diagonalization technique, which
provides us directly with the desired state without an explicit relaxation dynamics. 
In the counterdiabatic approach the same phenomenon is observed. \cite{DelCampo_2012}
The counterdiabatic approach might therefore be helpful for an experimental realization of our state tracking.

We apply this method within the framework of the density matrix renormalization group (DMRG) technique \cite{White_Noack_1992, *White_1992, *White_1993}.
However, our concept is not restricted to the DMRG and can be implemented in any wave function based technique.
As an example, we follow the current-carrying states of a 1D ring of interacting spinless fermions pierced by an Aharonov-Bohm flux. 
Moreover, in relation with the few recent studies devoted to the entanglement entropy of \emph{excited states} \cite{Masanes_2009, Alba_2009, Alcaraz_2011, *Berganza_2012, Eloy_2012}, we also show that our method enables the computation of the entanglement entropy of the tracked excited states.

\textit{The adiabatic tracking method}: 
DMRG \cite{White_Noack_1992, *White_1992, *White_1993} is a powerful technique to study the equilibrium properties of 1D interacting systems. 
It was extended to nonequilibrium situations, such as quantum quenches \cite{Roux_2010}, by the development of time-dependent simulations.\cite{Schmitteckert_2004, White_2004, *Daley_2004} 
Standard DMRG procedures can provide the GS wave function and a few (hundred) low-lying excited states in systems of a few hundred lattice sites. For higher excitations one has to resort to exact diagonalization techniques, which  are restricted to small systems. In this work we propose an approach to track excited states that cannot be reached within standard DMRG and we apply it to systems sizes that cannot be computed with exact diagonalization.

The system is described by a Hamiltonian $\mathcal{H}(\Phi)$, where $\Phi$ is a tunable parameter, such as an Aharonov-Bohm flux. As a start, a first DMRG run is performed with $\Phi=\Phi_0$ and determines the GS of the system $|\Psi(\Phi_0)\rangle$. 
Then the procedure restarts with a new value of the parameter $\Phi_1 = \Phi_0 + \delta \Phi$ but this time it searches for an eigenstate that maximizes $\langle\Psi(\Phi_0)| \Psi(\Phi_1)\rangle$, the overlap with the previous eigenstate. 
In the results reported below, this overlap is in general very close to one ($\gtrsim 0.99$) and never falls below 0.72.
The previous eigenstate is kept using the wave function prediction technique \cite{White:PRL96},
which is also at the heart of the adaptive time evolution schemes.\cite{White_2004,*Daley_2004} 
In other terms, instead of keeping the GS, as in usual DMRG procedures, it determines the state that is the most similar with the previous one and further computations are performed with that state. Increasing slowly the value of the parameter $\Phi$ up to a final value $\phimax$ thus leads the system each time higher in its spectrum and allows the tracking of a stationary excited state. 
Once $| \Psi(\phimax)\rangle$ is reached, one can study the effect of another perturbation on this excited state.
A second perturbation is switched on, e.g. a Coulomb interaction tuned by a parameter $U$, and the procedure restarts from  $| \Psi(\phimax,U_0)\rangle$ up to $| \Psi(\phimax,\Umax)\rangle$.
It should be stressed that if the first perturbation opens a gap the method will just follow the GS, as in the usual Landau-Zener scenario (Fig.~\ref{fig:cartoon}).
\begin{figure}[!b]
\includegraphics[width=0.49\textwidth]{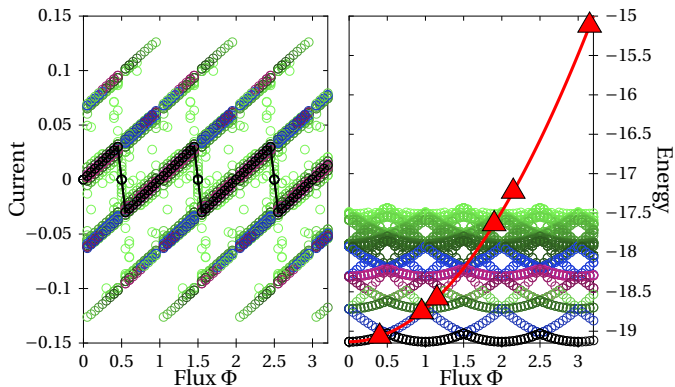} 
\caption{\label{fig:50_levels} $50$ lowest energy levels and currents of Eq.~(\ref{eq:AB_term}) versus the flux $\Phi$ computed with a standard DMRG procedure in a $L=30$ ring. Different colors correspond to different levels. 
Left panel: the current of the GS is highlighted with a black line. 
Right panel: the red curve shows the tracked states and the triangles correspond to values discussed in the text ($\phimax=0.4,0.95, 1.15, 1.9$, $2.15$ and $3.15$). }
\end{figure}
In order to access excited states we first follow the system through level crossings of the clean system and
turn then interaction on, which may lead to avoided level crossings for smaller flux values.
As a final note we would like to remark that from a technical point of view this state evolution scheme combines
the advantage of the adaptive time evolution schemes \cite{White_2004,*Daley_2004} of having only two states
to track in each DMRG run with the advantage of the full time evolution scheme \cite{Schmitteckert_2004} where
one always starts from an initial state based on a diagonalization technique.

\begin{table}[htb]
 {
 \begin{tabular}{|r*{9}{|c}|}
  \hline\hline
  $\Phi$                           & 0.4 & 0.9 & 1.15 & 1.6 &  1.9 & 2.15 &  2.6 & 2.9  & 3.15  \\[1.0ex]
  \hline 
  L=30                             &   0 &    1 &    4 & 17 & 36 &   88 &  451 & 1346 & 3289 \\[1.0ex]
  \hline 
  L=34                             &   0 &    1 &    4 & 17 & 34 &   87 &  441 & 1276 &  3187 \\[1.0ex]
   \hline 
  L=38                             &   0 &    1 &    4 & 17 & 34 &   85 &  415 & 1223 &  3067 \\[1.0ex]
  \hline 
   L=42                            &   0 &    1 &    4 & 17 & 34 &   83 &  389 & 1173 &  2905 \\[1.0ex]
   \hline  
   L=50                            &   0 &    1 &    4 & 17 & 34  &  79 &      &      &       \\[1.0ex]
  \hline
  \end{tabular}
 \begin{tabular}{|r*{6}{|c}|}
 \hline
  $\Phi$                           &   3.6      & 3.9    & 4.15 &  4.6 &   5.15   &    6.15   \\[1.0ex]
  \hline 
  L=30                             &  15283     & 41381  & 91 996 &  356 484 &   1 582 279   &   13 769 513   \\[1.0ex]
  \hline\hline
 \end{tabular}
 }
\caption{\label{tab:LevelIndex} Eigenstate index $n$ of an adiabatically tracked state vs.\ the eigenstate index obtained from an exact diagonalization 
         of the quadratic form for system sizes of $L=30$, 34, 38, 42 and 50 sites.}
\end{table}
\textit{Current-carrying state in a ring of spinless fermions}:
We now put into practice the adiabatic tracking method with a simple model. We consider a 1D ring of size $L$ of spinless fermions pierced by an Aharonov-Bohm flux $\Phi$:
\bea
\mathcal{H} &=& - t \sum_{x=1}^L e^{\im 2\pi\Phi/L}  \hat{c}_{x-1}^{\dagger} \hat{c}_{x}^{} + \rm{h.c.}
	    \label{eq:AB_term}	 
	    \\
            &+& U \sum_{x=1}^L\big(\hat{n}_{x-1} - \frac{1}{2}\big)\big(\hat{n}_{x} - \frac{1}{2}\big)
	    \label{eq:interaction}
\eea
$\hat{n}_x=\hat{c}_x^{\dagger} \hat{c}^{}_x$ is the density operator  and the flux $\Phi$ is in units $\Phi_0= hc/e$. In this work we focus on half-filled ($n=1/2$) systems. Different fillings would affect the parity of the persistent currents but not the method discussed hereafter. Umklapp scattering is prevented by considering only incommensurate values of $\Phi$. 
In the absence of the flux, model (\ref{eq:AB_term},\ref{eq:interaction}) maps to the integrable $XXZ$ model \cite{Yang_Yang} through the Jordan-Wigner transformations. In the weakly interacting region it is described by a Gaussian theory, a Luttinger Liquid (LL) \cite{Haldane_1980, *Haldane_1981}, up to $U_c=2$ where it undergoes an Ising-like phase transition to a gapful long range ordered N\'eel state. The renormalization group equations, and thus $U_c$, are not affected by the presence of the flux \cite{Loss_1992}. 
The latter simply shifts the values of the particle momenta $p \rightarrow p + \Phi/L$ and the excitation spectrum remains compact.

We use a DMRG algorithm to compute the lowest energy levels and the expectation value of the current operator 
$\hat{J}_x = -4\pi t \cdot\mathrm{Im} \big[ e^{\im 2\pi\Phi/L} \langle c_{x-1}^{\dagger}c^{}_{x} \rangle \big]$. 
Let us first start with the noninteracting case Eq.~(\ref{eq:AB_term}).  
The energy spectrum consists in a set of flux-periodic energy levels. 
Fig.~\ref{fig:50_levels} shows the $50$ lowest levels, obtained with a standard DMRG procedure in a $L=30$ ring, together with the corresponding persistent currents.   
Here we may stress that keeping so many levels, even in a noninteracting case, is already involved. 
Indeed, targeting more than a few low-lying levels requires computing more eigenvalues and eigenstates during the diagonalization of the superblock Hamiltonian. 
The numerical cost increases since this requires a very large number of Davidson iterations, exceeding a few times the number of low-lying states kept (see for example Ref.~\onlinecite{Manmana_2005}).
In addition one has to increase the target space to provide a faithful representation of all the desired states. Yet one is facing the problem that DMRG in general provides
accurate eigenstates, however it is very hard to ensure that indeed all desired eigenstates are found as it may happen that
some states of the spectrum are missing. Therefore special care has to be taken in order to ensure finding the complete low energy spectrum,
e.g.\ see \cite{Rapp:NJP13}.
\begin{figure}[!b]
\includegraphics[width=0.49\textwidth]{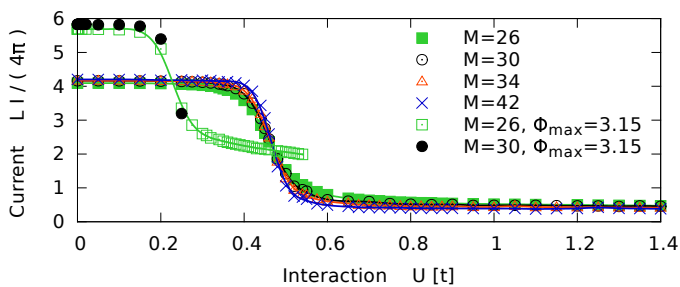} 
\caption{\label{fig:E_I_versus_U} Current versus interaction for excited states taken at $\phimax=2.15$ and $3.15$ in rings of various lengths.}
\end{figure}
In order to overcome these problems and to reach higher excitations disregarding the limitations of the standard DMRG procedure, we now turn to our adiabatic tracking method. 
We performed DMRG with periodic boundary conditions, performing between 7 and 11 sweeps and keeping up to 5000 states per block. 
The discarded entropy ranges between $10^{-11}$ and $10^{-4}$.
We considered rings up to $50$ sites of noninteracting spinless fermions and report results for an applied a flux up to $\phimax=2.15$ and $3.15$ (see red triangles in Fig.~\ref{fig:50_levels}). 
For the noninteracting systems we checked our DMRG results by comparing to an exact diagonalization of the quadratic form.
By looking at all possible particle hole excitations with energies below or equal the energy of the tracked states we could identify all tracked states.
In table~\ref{tab:LevelIndex} we provide the eigenenergy index for the states we tracked adiabatically within DMRG, where index '0' denotes the GS.

Then the nearest-neighbor Coulomb interaction (\ref{eq:interaction}) is added.
Fig.~\ref{fig:E_I_versus_U} shows the energy and current of $| \Psi(\phimax,U)\rangle$. The magnitude of the current depends on $1/L$; for the GS, $J^{\rm GS}_{U=0} \sim \frac{\mathcal{D}\Phi}{L}$, where $\mathcal{D}$ is the charge stiffness which depends on the Fermi velocity and Luttinger $K$ parameter \cite{Giamarchi}. 
Because of extra operators that become relevant for excited states, the region before the current drop becomes smaller as $\phimax$ increases.
We assume that this drop is related to the formation of bound states in the Bethe ansatz solution, see \cite{Woynarovich_1987, *Hamer_1987}.
We point out that the ordering of the current with respect to system size changes at the drop, so this effect persists in the thermodynamic limit.
\begin{figure}[!b]
\includegraphics[width=0.48\textwidth]{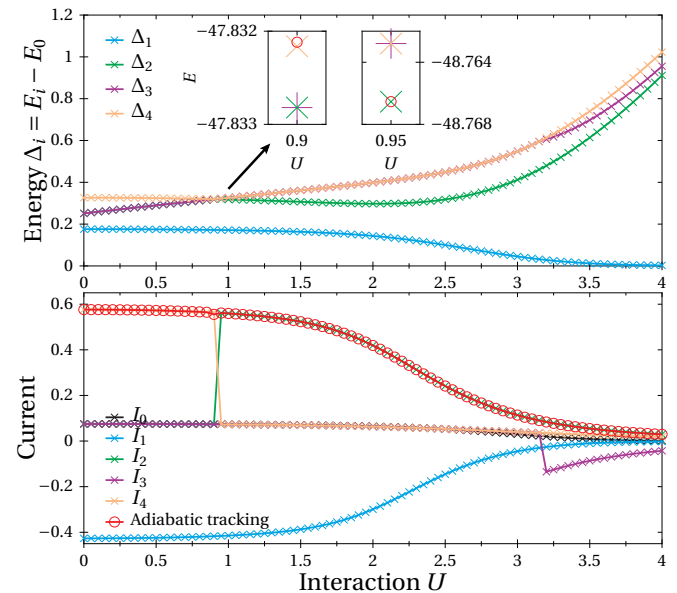} 
\caption{\label{fig:Level_Tracked} Comparison between the adiabatic tracking (red dots) and standard DMRG (crosses) keeping the $5$ lowest levels at $\phimax=1.15$ in $L=50$ rings. Top panel: energy differences between excited states and the GS $\Delta_i=E_i-E_0$. Inset: the adiabatic tracking (red dot) comes trough the energy level crossing between the $4^{\rm th}$ and the $2^{\rm nd}$ levels. Bottom panel: currents of the corresponding states.}
\end{figure}

In order to gain more information on the nature of the excited state $| \Psi(\phimax,U)\rangle$, and also to show that it is not lost at an energy level crossing, we compare data obtained with the adiabatic tracking method and with standard DMRG. We keep five low-lying levels at $\phimax=1.15$, where the comparison is simple (see Fig.~\ref{fig:50_levels}), see also \cite{Gendiar_2009}.
Fig.~\ref{fig:Level_Tracked}, top panel, shows the energy differences between the $n^{\rm th}$ level and the GS. It appears that increasing $\Phi$ up to $\phimax=1.15$ brings the system to its $4^{\rm th}$ excited state, which is consistent with the localization of the corresponding triangle in  Fig.~\ref{fig:50_levels}. As the interaction increases, an energy level crossing takes place around $U \sim 0.92$ and the tracking continues with the $2^{\rm nd}$ excited state. The inset shows that the incrementation of the interaction $\delta U=0.05$ is small enough so that the state tracked is not lost at the energy level crossing.
The current (bottom panel) of the $4^{\rm th}$ level undergoes a jump at the energy level crossing and crosses the of the $2^{\rm nd}$ level. One observes that, as expected, the tracking method keeps on with the current-carrying state. 
This actually reflects the adiabatic character of out state evolution: the observables change smoothly.
We have also checked that the adiabatic tracking for $\phimax < 1/2$ (i.e. still in the GS) reproduces the Bethe ansatz exact solution for the current in the critical region \cite{Woynarovich_1987, *Hamer_1987}.

\textit{Entanglement entropy of excited states}: 
DMRG algorithms are based on the computation of the reduced density matrix $\rho_A$ of a subblock $A$ with length $\ell$ obtained by tracing over the degrees of freedom of the complementary block $\bar{A}$,  $\rho_A=\Tr_{\bar{A}}[\rho]$. The entanglement entropy is the corresponding von Neumann entropy: $S_A= -\Tr [\rho_A \ln \rho_A$ ]. 
One of the most important results for gapped systems with short range interactions is that $S_A$ satisfies an area law \cite{Vidal_2003, *Latorre_2004} and is proportional to the hyper-surface $\ell^{d-1}$ separating $A$ from $\bar{A}$. 1D systems ($d=1$) are thus particular since $S_A^{\rm GS}$ is independent from $\ell$ and bounded by a constant that depends on the width of the gap between the GS and the first excited state.\cite{Hastings_2007, *Eisert_2010} 
\begin{figure}[!b]
\includegraphics[width=0.49\textwidth]{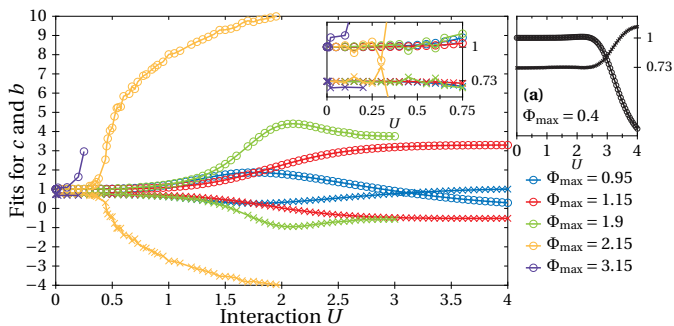} 
\caption{\label{fig:CFT} Fits of $c$ and $b$ from Eq.~\ref{eq:entropy_scaling} versus interaction at different values of $\phimax$ (see Fig.~\ref{fig:50_levels}) in $L=50$ rings, excepted $L=30$ for $\phimax \ge 1.90$. Right (a) panel: $\phimax=0.4$, the system is still in its GS. The errors bars are not showed since they are smaller than the symbols.}
\end{figure}
Critical systems are known to violate this area law. For $(1+1)-$dimensional systems with conformal invariance, such as the $XXZ$ model, the entanglement entropy increases logarithmically with $\ell$ and is proportional to the central charge $c$: \cite{Holzhey_1994, Calabrese_2009}.
\be
 \mathcal{S}_A^{\rm GS}(\ell) = \frac{c}{3} \ln\big(\frac{L}{\pi}\sin\frac{\pi \ell}{L}\big) + b
 \label{eq:entropy_scaling}
\ee
where $b$ is a nonuniversal constant, whose analytical expression is known exactly for the $XXZ$ model \cite{Jin_Korepin}. Eq.~(\ref{eq:entropy_scaling}) is valid for periodic boundary conditions.
Fig.~\ref{fig:CFT} (a) shows the fits obtained for a ring of 50 sites at $\phimax=0.4<1/2$ (lowest triangle of Fig.~\ref{fig:50_levels}). We recover $c=1$ in the critical region $U<2$ and $b \sim 0.73$ in agreement with the predicted value. For $U>2$, $S^{\rm GS}_A \approx b$ agrees with the area law.

The adiabatic tracking method gives access to the entanglement entropy of the tracked \emph{excited state}. 
Eq.~(\ref{eq:entropy_scaling}) stills applies in the critical region and $c$ is then interpreted as the central charge of an effective Hamiltonian whose GS is our tracked excited state.\cite{Alba_2009} 
Fig.~\ref{fig:CFT} shows fits of (\ref{eq:entropy_scaling}) versus interaction at the values of $\phimax$ of Fig.~\ref{fig:50_levels}. In order to understand their behaviors, we turn to the few recent works devoted to the exact calculation of the entanglement entropies of excited states, $S_A^{\rm exc}$.\cite{Masanes_2009, Alba_2009, Alcaraz_2011, *Berganza_2012}  
For excited states engendered by the action of a primary field on the GS, $S_A^{\rm exc}$ differs from $S_A^{\rm GS}$.\cite{Alcaraz_2011, *Berganza_2012} It can still be computed by means of a scaling function, which, for the vertex operators of the $XXZ$ model, turns out to be equal to one in the thermodynamic limit. More generally, excited states generated by compact excitations (i.e. that engender no holes in the spectrum) should have $S_A^{\rm exc} = S_A^{\rm GS}$. 
We indeed find $c=1$ and $b \sim 0.73$ in the critical region for all the values of $\phimax$ considered (see inset in Fig.~\ref{fig:CFT}). 

Away from criticality, it is difficult to give a general picture. 
As a matter of fact, Fig.~\ref{fig:CFT} shows radically different behaviors for different values of $\phimax$. 
One could expect an excited state separated by a gap from the next excited states to obey some type of area law.\footnote{Ref.~\cite{Masanes_2009} actually provides criteria, e.g. on the behavior of the correlations functions, to determine if low-lying excited states obey an area law.} The fit obtained at $\phimax=0.95$ with $S^{\rm exc}_A \approx b$ at large $U$ suggests such behavior. 
However, for other $\phimax$, $S^{\rm exc}_A$ evolves from one region with logarithmic divergence (the critical region) to another at larger $U$. We observe a saturation of $c$ and $b$ (e.g. $c \sim 3.2$ for $\phimax=1.15$). 
As stated in Ref.~\cite{Alba_2009}, the value of $c$ increases with the number of discontinuities in the spectrum. 
However, a full understanding of this behavior is beyond the scope of this work. Intensive simulations together with the study of the correlation functions may be required if one wanted to study in details these values.

Finally, we would like to remark that we are not restricted to track the evolution of the GS.
We could also take another low lying state obtained by a direct calculation as a starting state 
and follow its evolution. Such an extension would give access to different kind of excited states.

\textit{Conclusion}: We have proposed a new method to track the excited state reached by a quantum system after a perturbation that is slowly switched-on. 
We have shown using DMRG that this adiabatic tracking follows accurately the current-carrying states of a ring of spinless fermions under an Aharonov-Bohm flux.
Of course one does not have to follow the protocol of the external perturbation used in this manuscript.
The method is very flexible and allows to track any perturbation that can be implemented.
In addition, we can extract the entanglement entropy of excited states in large systems and without being limited to integrable models. 
An example of further application is the computation of the Luttinger $K$ parameter which can be extracted from the number fluctuation in the subblock $A$ \cite{Rachel_LeHur_2010, *Nishimoto_2011} or from the oscillations of the R\'enyi entropies for $n\ge2$ \cite{Calabrese_Essler_2010}, which may obey the usual GS scaling also for excited states.\cite{Alba_2009}

Finally, the main difference between our adiabatic state tracking and evolution schemes is that we are tracking eigenstates of the systems, while time dependent
quenches are in general not following eigenstates. Therefore, the adiabatic state tracking
resembles the concept of the adiabatic switching on of perturbations as applied in scattering theory.
By directly following the eigenstates we are not restricted to systems where the scattering perturbation
has only a small impact on the bulk system.  It will interesting for future research to
investigate whether this concept can be used to extend the Lippmann-Schwinger type calculations
to finite systems.
\begin{acknowledgments}
We thank P. Azaria for insightful comments.
\end{acknowledgments}

\bibliographystyle{apsrev4-1}

%

\end{document}